\documentstyle[prb,twocolumn,aps,psfig,amssymb,floats]{revtex}

\begin{document}
\draft

\twocolumn[\hsize\textwidth\columnwidth\hsize\csname @twocolumnfalse\endcsname


\title{Semiclassical action based on dynamical mean-field theory
describing electrons interacting with local lattice fluctuations}
\author{Stefan Blawid and Gertrud~Zwicknagl} 
\address{Institut f\"ur Mathematische Physik, 
TU Braunschweig, Mendelssohnstr. 3, 38106 Braunschweig}

\date{\today}

\maketitle


\begin{abstract}
We extend a recently introduced semiclassical approach to calculating
the influence of local lattice fluctuations on electronic properties
of metals and metallic molecular crystals. The effective action of
electrons in degenerate orbital states coupling to Jahn-Teller
distortions is derived, employing dynamical mean-field theory and
adiabatic expansions. We improve on previous numerical treatments of
the semiclassical action and present for the simplifying Holstein
model results for the finite temperature optical conductivity at
electron-phonon coupling strengths from weak to strong. Significant
transfer of spectral weight from high to low frequencies is obtained
on isotope substitution in the Fermi-liquid to polaron crossover
regime.
\end{abstract}



\vskip2pc]


\section{Introduction}
\label{sec:intro}

Despite many years of research, techniques to calculate the influence
of strong vibronic interactions on electronic properties of metals and
metallic molecular crystals are urgently needed. Strong couplings of
conduction electrons to local lattice fluctuations have been observed
in certain phonon mediated superconductors with high transition
temperatures and manifest itselves among other effects in unexpected
mid-infrared transitions in the optical conductivity in these
materials, like in $\rm Ba_{1-x}K_xBiO_3$\cite{puchkov96} and in the
alkali-doped $\rm C_{60}$ fullerides.\cite{degiorgi94,gunnarsson97}
Standard Migdal Eliashberg (ME) theory\cite{migdal58} is not
sufficient to explain these phenomena although typical phonon
frequencies of physical interest may remain small compared to typical
electron energies. The reason for the failure is that ME theory
assumes that the underlying electronic ground state can be described
by Fermi liquid theory. However, if the electron-lattice coupling
strength $\lambda = \Lambda/t$ exceeds a critical value (of order
unity) the conduction electrons are believed to form ``small
polarons'',\cite{alexandrov94} so that the electronic groundstate is
fundamentally reconstructed.

Signatures of polaronic behavior have been most accurately studied in
the case of one or few electrons employing the Lang-Firsov
transformation.\cite{lang63} However, its applicability to the
physically relevant case of metallic densities is not established. A
suitable method to study the interaction of conduction electrons with
{\it local} lattice fluctuations may be the dynamical mean-field
theory\cite{georges96} and in recent years an increasing number of
studies have been published. In spite of the substantial
simplifications which occur when the conduction electrons are
described by local fluctuating fields the resulting equations are
still very complicated. An exact solution is only possible in the case
of a single polaron.\cite{ciuchi96} Numerically ``exact'' solutions
are avaiable using quantum Monte Carlo methods\cite{freericks} in the
antiadiabatic limit $\Omega \sim t$ of unclear physical relevance and
using numerical renormalization techniques\cite{meyer02} at zero
temperature. Especially the latter is a promising approach at least
for simplifying models like the Holstein model and may be extended in
future to finite temperatures. Semianalytical techniques have the
advantage of being instructive and of being extensible to more
realistc model systems of conduction electrons coupling to local
lattice distortions including maybe even electron-electron
interactions. These
approaches\cite{millis96,benedetti98,deppeler00,blawid02} rely on a
detailed analysis of the effective action based on the dynamical
mean-field theory.

Here, we extend a ``semiclassical'' approach\cite{blawid02} recently
introduced by one of the authors to general models of electrons in
degenerate orbital states coupling to Jahn-Teller distortions. The
classical (static) phonon modes are treated exactly and the quantum
(dynamical) modes are expanded to second order. The adiabatic
expansion of the classical action allows for treating polaronic
effects in the physical relevant limit of small adiabatic parameter,
from weak to strong electron-phonon couplings. However, the method is
restricted to temperatures which exceed or are comparable to an energy
scale set by a renormalized phonon frequency. Numerically, the method
is implemented for the simplifying Holstein model of (spinless)
electrons at half filling. We resolve previous numerical
restrictions\cite{blawid02} connected with the analytic continuation
of the local electronic Green function from the imaginary to the real
frequency axis which allows for a more complete understanding of
dynamical quantities, especially the optical conductivity. A treatment
of coupling strengths beyond the classical polaronic instability
becomes numerically tractable.  Pioneering work in the calculation of
the density of states and the optical conductivity at low temperatures
has been done by Benedetti and Zeyher\cite{benedetti98} and we provide
a comparison to their results. The remainder of the paper is organized
as follows. In Sec. \ref{modelling} we apply the dynamical mean-field
theory to models of metallic molecular crystals or metals with strong
Jahn-Teller distortions. The polaronic instability for classical
vibrations is discussed in Sec. \ref{classical} and known results in
the opposite quantum limit at zero temperature are reviewed in
Sec. \ref{quantum}. Section \ref{adiabatic} gives the adiabatic
expansion of the effective action and discusses the numerical
implementation to treat the Holstein model at half-filling, as well as
the limitations of the approach. Numerical results for the density of
states and the optical conductivity are presented and discussed
extensively. We conclude in Sec. \ref{conclusion}.

\section{Modelling}
\label{modelling}

We study metallic molecular crystals whose Hamiltonians are described
by
\begin{equation}
\label{hamil}
H = \sum_i\,H_i^{\rm Molecule} + 
\sum_{i,j,\alpha,\beta}\,t_{ij}^{\alpha\beta}\,
c_{i\alpha}^{\dagger} c_{j\beta}^{\phantom{\dagger}}
-\mu\,\sum_{i,\alpha}\,
c_{i,\alpha}^{\dagger} c_{i,\alpha}^{\phantom{\dagger}} 
\;.
\end{equation}
Here, the operator $c_{i\alpha}^{\dagger}$ creates an electron in the
molecular orbital (MO) $\alpha$ on the $i$-th molecule. For simplicity
we suppress the spin index and consider spinless electrons. However,
our treatment can be easily extended to a paramagnetic state by
including the appropriate spin degeneracy.  We will assume that the
molecular orbital state are single particle like, e.g.~provided by a
Hartree-Fock treatment of the molecule. In particular, the hopping of
electrons on and off a molecule shall not alter the local electronic
level scheme. In this respect the assumption of spinless electrons
helps to avoid unphysical large fluctuations in the local occupation
number of the partially filled MOs giving rise to the conduction
bands.

In general, the high local symmetry of a molecule will lead to an
orbital degeneracy of the interesting, partially filled MOs. Therefore
$H_{\rm Molecule}$ describes the coupling between a $f$-fold
degenerate electronic term and $N$ Jahn-Teller active modes
\begin{eqnarray}
H_{\rm Molecule} & = & \frac{1}{2}\,
\sum_{i=1}^{N}\,(K Q_i^2+ M \dot{Q}_i^2) \\ 
& & +\left(
\begin{array}{ccc}
W_{11}(Q_1,\ldots,Q_N) & \cdots & W_{1f} \\
\vdots & \ddots & \vdots \\
W_{f1} & \cdots & W_{ff} 
\nonumber
\end{array}
\right)\;.
\end{eqnarray}
Here, the $Q_i$ denote the normal coordinates of the Jahn-Teller
distortions. According to the Jahn-Teller theorem the leading term of
the coupling constants $W_{ij}$ is linear in $Q$ and different matrix
elements are symmetry related. Employing group theory we may
characterize different cases according to different irreducible
representations of all possible point groups. In particular $E\otimes
e$ denotes the coupling of a double degenerated electronic term with
two Jahn-Teller modes.

We do not attempt to solve (\ref{hamil}) directly which is a very
difficult problem. Instead we are analysing an {\it effective action}
of the form 
\begin{eqnarray} 
\label{effective} 
S_{\rm eff} & = & \frac{1}{2T}\, 
\sum_{i=1}^N\,\sum_k Q_i(\omega_k) (K+M\omega_k^2)
Q_i(\omega_{-k}) - \mu\,n  \\ 
& & - {\rm Tr}_n \, {\rm Tr}_{\rm orb} \ln
\left\{ c(\nu_n)-\underline{\underline{W}}[\vec{Q}(\nu_n-\nu_m)]
\right\} \;.
\nonumber  
\end{eqnarray} 
This is the action of a single molecule embedded in an effective
environment when the fermionic fields (which occur in quadratic form)
are integrated out. The latter give rise to the mean-fields $c(\nu_n)$
which describe the conduction electrons and depend on odd Matsubara
frequencies $\nu_n = 2 \pi T (n+1/2)$. The $Q_i(\omega_k)$ are bosonic
fields (describing the Jahn-Teller modes) depending on even Matsubara
frequencies $\omega_k = 2 \pi T k$. Note that the
$\underline{\underline{W}}$-term in the logarithm is nondiagonal in
both the Matsubara and the orbital index.  The action $S_{\rm eff}$
arises in a {\it dynamical mean-field} (DMFT) \cite{georges96}
description of the original lattice model Eq. (\ref{hamil}) if we
assume a structureless and fast fluctuating environment of a single
molecule, i.e.~if the local Green function can be averaged over the
$f=n_{\rm orb}$ orbital degrees of freedom
\begin{equation} 
\label{average} 
{\cal G} = \frac{1}{n_{\rm orb}}\,{\rm Tr_{\rm orb}}\; 
\underline{\underline{G}}_{\rm loc}\;.
\end{equation} 
The orbital averaging was first introduced in
Ref. \onlinecite{millis96}. The DMFT treatment provides a
self-consistency expression for the dynamic mean-fields $c$. The
partition function may be written as a functional integral over the
bosonic fields
\begin{equation} 
\label{partition} 
Z = \int {\cal D}[Q]\exp(-S_{\rm eff}).  
\end{equation} 
from which the local Green function ${\cal G}$
follows 
\begin{equation} 
\label{greens} 
{\cal G}(\nu_n) \equiv \frac{\delta\,\ln Z}{\delta\,c(\nu_n)}\;.  
\end{equation} 
The molecules are arranged on a three dimensional lattice. However,
after averaging over the orbital degrees of freedom in (\ref{average})
it may be sufficient to work with an effective density of states of
simple form, e.g.~semicircular $\rho_0(\epsilon
)=(1/N)\,\sum_{\vec{k}}\delta (\epsilon-\epsilon_{ \vec{k}}) = 1/(2\pi
t^2)\sqrt{4 t^{2}-\epsilon ^{2}}$. In this case the self-consistency
equation (\ref{average}) can be written in the form
\begin{equation}
\label{dmft} 
c_n = i\omega _{n}+\mu -\frac{t^2}{n_{\rm orb}}\, 
{\cal G}_n \left( \{c_n\} \right)\;.  
\end{equation} 
Eqs.~(\ref{effective}), (\ref{partition}), (\ref{greens}) and
(\ref{dmft}) form a complete set of equations which, in principle, can
be solved for the orbital averaged local Green function on the
imaginary axis of a molecular crystal with Jahn-Teller active local
distortions. However, one cannot evaluate the path integrals exactly
in Eq. (\ref{partition}) and further approximations are necessary. We
adopt the adiabatic expansion of the effective action Eq.
(\ref{effective}) which includes the leading term in the non diagonal
part of the electron-phonon interaction as discussed in
Sec. \ref{adiabatic}.

The present modelling is not restricted to the description of metallic
molecular crystals as pioneering works on the ``colossal
magnetoresistance'' manganites demonstrate.\cite{millis96} However,
when applying the present formalism to others than molecular crystals
one have to keep in mind that local distortions of the crystal unit
cell do influence the neighboring cells. In particular, in transition
metal oxides $E_g$ symmetry distortions of the oxygen octahedra move
oxygen atoms of several unit cells. Therefore, in this case the
presented modelling is only applicable to phenomena which involve
incoherent features of the electronic Green function, i.e.~which
involve high temperatures or high frequencies.

\section{Classical action: Analytical insights}
\label{classical}

A good starting point to gain some insight in the physics described by
the Eqs. (\ref{effective}) - (\ref{dmft}) is to analyse the effective
action Eq. (\ref{effective}) in the high temperature
limit\cite{millis96} when only thermal fluctuations are present.
$S_{\rm eff}$ reads
\begin{eqnarray}
S_{\rm class}(\vec{Q}) & = & \frac{K}{2T}\,\sum_{i=1}^N\,Q_i^2-
\mu\,n \\
& & - {\rm Tr}_n \, {\rm Tr}_{\rm orb} \ln
\left\{
c(\nu_n)-\underline{\underline{W}}[\vec{Q}]
\right\}
\nonumber
\end{eqnarray}
where the $Q_i=Q_i(\omega_k = 0)$ are the classical components of the
bosonic fields. The function
\begin{equation}
\label{prob}
P(\vec{Q}) = \exp [-S_{\rm class}(\vec{Q})]
\end{equation}
is the (unrenormalized) probability that the classic Jahn-Teller
distortion takes the value $\vec{Q}$.

\begin{figure}
\centerline{\psfig{file=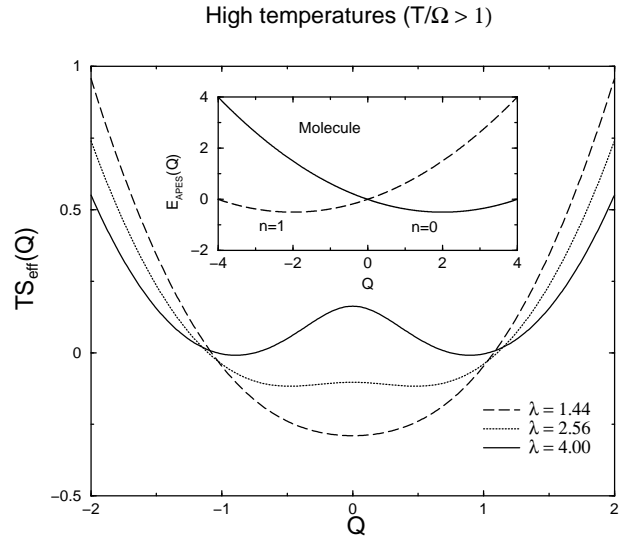,width=8cm,angle=-90}}
\vspace*{0.1cm}
\caption{\label{high.fig} \footnotesize MAIN PANEL: The
(classical) effective action $S_{\rm eff}$ as function of the
distortion $Q$ for the Holstein model at half-filling. Shown is the
behavior for different coupling strengths $\lambda$ at a high
temperature of $T/t=0.2$. The forming of a double well at a finite
value of $\lambda$ is clearly seen. INSET: Adiabatic potential energy
surface $E_{\rm APES}(Q) = Q^2/2\Lambda + Q(n-1/2)$ at $\lambda =
2.56$ for a molecule with no and one electron in the relevant molecule
orbital. The distortion equals $\pm\lambda$.}
\end{figure}
We specialize now to different cases. The case of a non degenerate
electronic term coupling to a totally symmetric distortion (Holstein
model) is the simpliest situation with the interaction matrix being a
scalar $\underline{\underline{W}} = g\,Q$. The strength of the
electron-phonon coupling is measured by $\Lambda = g^2/K$. (In the
following we will frequently denote the dimensionless coupling
parameter by $\lambda = \Lambda/t$). It is easy to show\cite{millis96}
that the effective action $S_{\rm class}(Q)$ changes from a single to
a double well behavior with increasing $\Lambda$ as shown in
Fig.\ref{high.fig}. In Fig.\ref{high.fig} the term $-Q/2$ has been
added to the effective action to symmetrize the distortions in the
case of hole and electron occupation on a given site. A double well
occurs when the $Q^2$-coefficient in the effective action
vanishes. This is the case for $\Lambda_c = -1/\Gamma_2$ with
\begin{equation}
\Gamma_2 = T\,\sum_n\,c^{-2}(\nu_n)\;.
\end{equation}
Here the fully self-consistent determined dynamical fields have to be
used. If we insert the free form, the critical electron-phonon coupling
reads\cite{deppeler02a}:
\begin{equation}
\Lambda_c = \frac{3\pi}{4}\,[1-(\mu/2t)^2]^{-3/2}\;.
\end{equation}
At couplings exceeding $\Lambda_c$ the motion of an electron is
accompanied by a lattice distortion, i.e.~small polarons are formed.
It is important to note that the behavior of an isolated molecule is
different. With a linear electron-phonon coupling a finite Jahn-Teller
distortion occurs for every finite coupling constant $\Lambda$ as
illustrated in the inset of Fig.\ref{high.fig}. In a crystal the
quantum fluctuations of the occupation numbers of the MOs due to the
hopping of the electrons smear out this distortion. A finite minimal
coupling strength is required for the lattice relaxing to the actual
occupation of a given site.

The occurence of finite Jahn-Teller distortions at a finite critical
value of $\Lambda$ is a generic feature of the classical action
$S_{\rm class}$. However, the critical value depends upon details of
the Jahn-Teller coupling, in particular upon the degeneracy of the
problem. This can be illustrated by comparing the $T_1\otimes
h$-problem describing the coupling between electrons and vibrations of
the charged $\rm C_{60}$ molecules to the $E\otimes e$-problem
relevant for transition metal oxides with perovskite
structure. Electrons added to an isolated fullerene molecule belong to
the $T_{1u}$ lowest unoccupied MO which can couple linearly to
vibrations of symmetry $h_g$.\cite{obrien97} The Jahn-Teller
interaction matrix reads
\begin{equation}
\underline{\underline{W}} = 
\frac{g}{2}\,\left(
\begin{array}{ccc}
Q_1-\sqrt{3}\,Q_4 & -\sqrt{3}\,Q_3 & -\sqrt{3}\,Q_2\\
-\sqrt{3}\,Q_3 & Q_1+\sqrt{3}\,Q_4 & -\sqrt{3}\,Q_5\\
-\sqrt{3}\,Q_2 & -\sqrt{3}\,Q_5 & -2\,Q_1
\end{array}
\right)\;.
\end{equation}
When introducing a new parametrization and expressing $Q_1,\ldots.Q_5$
via one amplitude $Q$ and four angles\cite{obrien97} the classical
effective action can be written in compact form
\begin{equation}
S_{\rm eff} = \frac{Q^2}{2\Lambda T} -
{\rm Tr}_n \ln \left[
c^3(\nu_n) - \frac{3}{4}\,Q^2\,c(\nu_n) + \frac{Q^3}{4}\,\cos 3\alpha
\right]\;.
\end{equation}
Expanding $S_{\rm eff}$ for small $Q$ reveals that the
$Q^2$-coefficient of the classical action vanishes for
\begin{equation}
\Lambda_c = -\frac{2}{3}\,\frac{1}{\Gamma_2}\;.
\end{equation}

\begin{figure}
\centerline{
\psfig{file=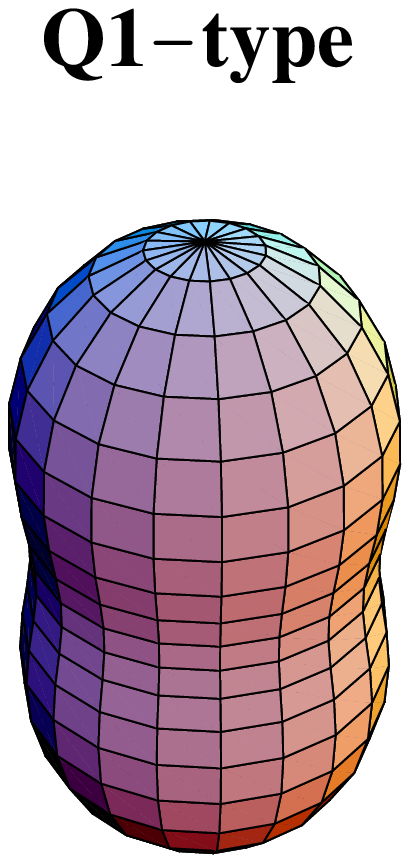,width=1.5cm,angle=0}
\psfig{file=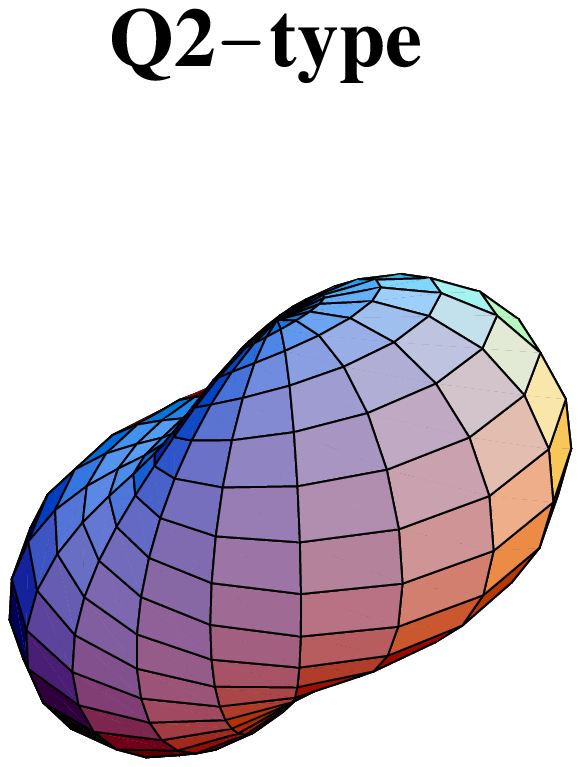,width=2cm,angle=0}
\psfig{file=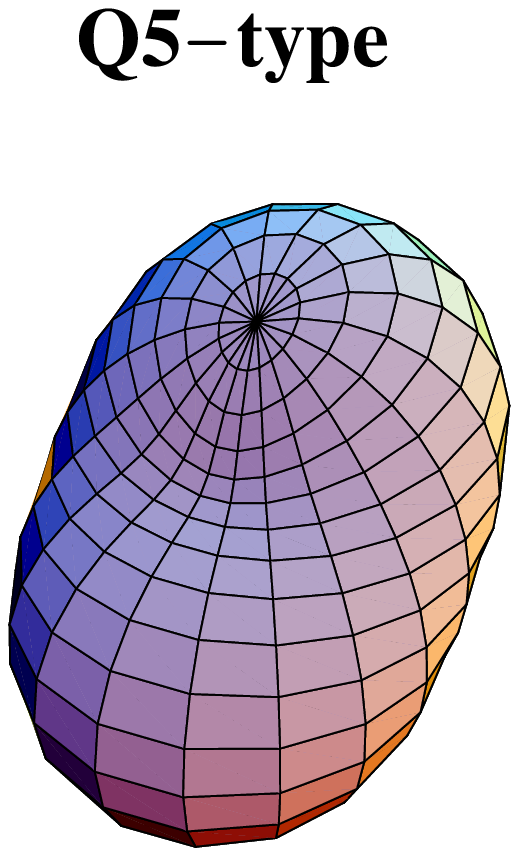,width=2cm,angle=0}
}
\centerline{
\psfig{file=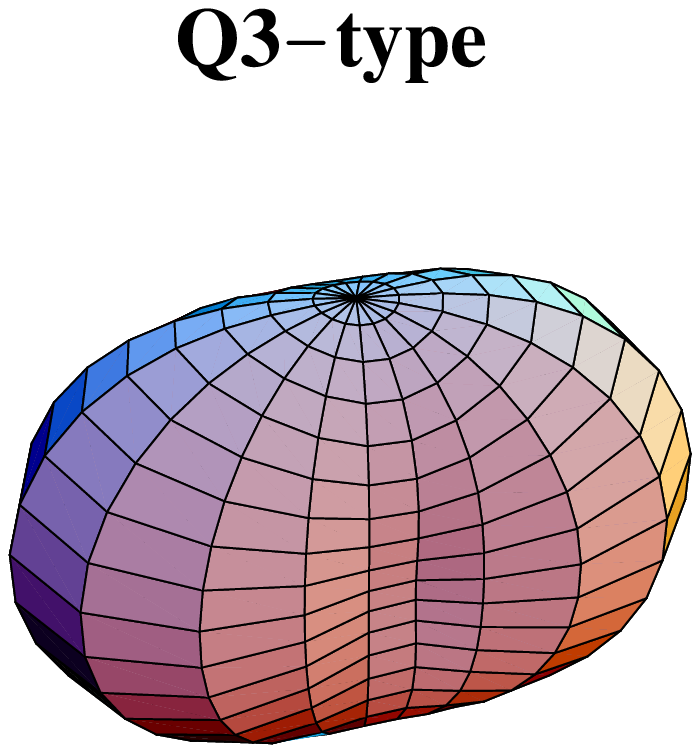,width=2cm,angle=0}
\psfig{file=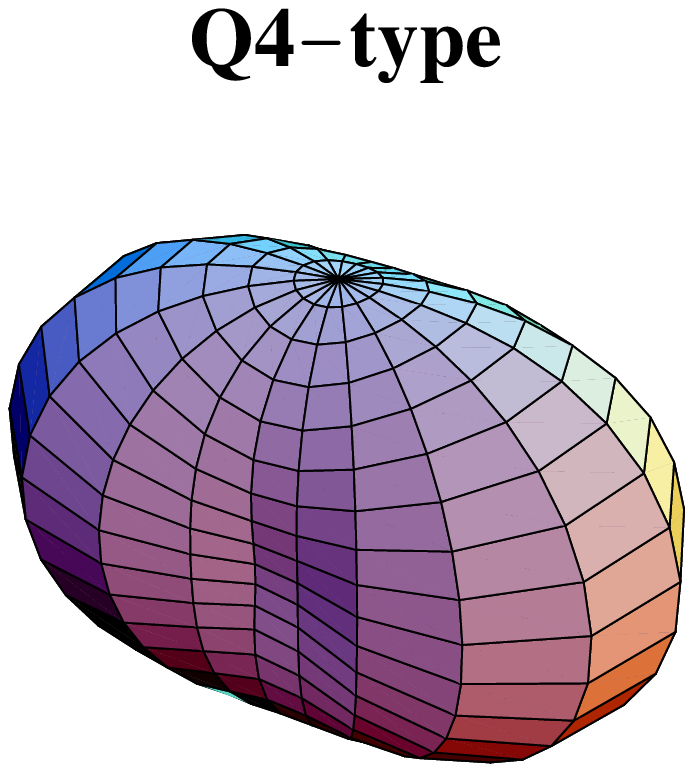,width=2cm,angle=0}
}
\vspace*{0.1cm}
\caption{\label{distort.fig} \footnotesize The five $h$-type
Jahn-Teller distortions measured relative to a sphere.}
\end{figure}
For coupling strengths $\Lambda > \Lambda_c$ finite Jahn-Teller
distortions occur as displayed in Fig. \ref{distort.fig}. For the
$E\otimes e$-problem one obtains\cite{millis96} $\Lambda_c =
-1/2\Gamma_2$. Wee see that for comparable values of the ``electron
bubbles'' $\Gamma_2$ the occurence of finite Jahn-Teller distortions
require stronger electron-phonon couplings in the higher degenerate
case. In this sense the weak-coupling regime for the $T_1\otimes
h$-problem extends to larger coupling strengths.

It is important to note that the occurence of a polaronic instability
indicates a break down of Migdal-Eliashberg theory because the ground
state is fundamentally reconstructed. Physically speaking a
quasiparticle picture does not hold in the transition regime because
neither a description as electrons nor as polarons is valid. This give
rise to interesting isotope effects on electronic properties as
discussed in Ref. \onlinecite{blawid02} and in Sec. \ref{adiabatic}.

\section{Influence of quantum fluctuations}
\label{quantum}

\begin{figure}
\centerline{\psfig{file=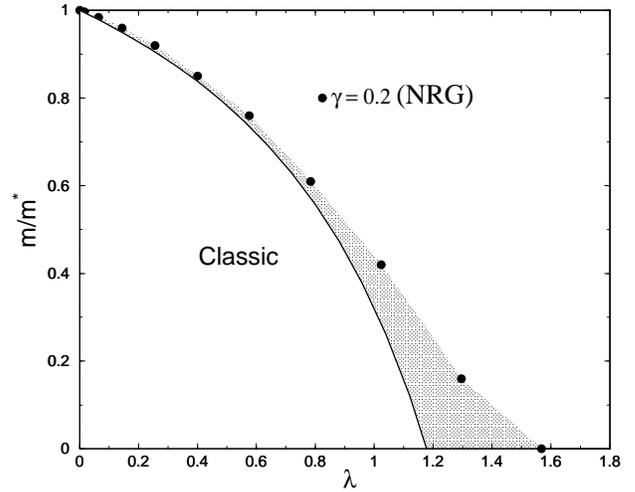,width=8cm,angle=-90}}
\vspace*{0.1cm}
\caption{\label{mass.fig} \footnotesize Inverse ratio of the effective
and bare electronic mass as function of $\lambda$. The classical
prediction is compared to the results obtained from an NRG study by
Meyer, Hewson and Bulla in Ref. 10 at finite phonon frequency. In the
shaded parameter regime strong isotope effects have to be expected.  }
\end{figure}
Quantum fluctuations of the lattice will change the value of the
coupling constant $\lambda_c$ needed to form small polarons. Even for
stronger electron phonon couplings when a classical lattice would
adjust to the occupancy of a given site, possible tunneling may hinder
the formation of polarons or even prevent it. At sufficiently low
temperatures only zero-point quantum vibrations of the lattice are
present and the classical picture is not valid. Detailed studies
exist\cite{meyer02,benedetti98,deppeler02a} for the evolution of the
electronic effective mass in the Holstein model as function of the
coupling strength $\Lambda$ in the adiabatic regime. All results
indicate that the electronic effective mass diverges (or at least
increases very rapidly) at a {\it finite} value of the electron phonon
coupling even in the presence of zero-point fluctuations. In
Fig. \ref{mass.fig} we replot the results of a numerical
renormalization (NRG) study published in Ref. \onlinecite{meyer02}.
The inverse ratio of the effective and bare electronic mass as
function of $\Lambda$ ($=2\,g^2/\omega_0$ in the notations of
Ref. \onlinecite{meyer02}) is shown. For the adiabatic parameter
$\gamma = \Omega/t = 0.2$ ($t=0.25$) a critical value of
$\lambda_c(\gamma) \approx 1.57$ has been reported. The classical
approach predicts $m^*/m = 1+\bar{\Lambda}\,\rho(\mu)$ with
$\bar{\Lambda} = \Lambda/(1-\Lambda/\Lambda_c)$, i.e.~a diverging
electron mass at the classical value $\Lambda_c$ which reads for a
semicircular density of states at half filling $\Lambda_c = 3 \pi
t/[(2s+1)4]$. The classical result is also shown in
Fig. \ref{mass.fig}. Note that in the NRG investigations the case of
twofold spin degenracy is considered, i.e.~the classical polaronic
instability occurs at $\lambda_c \approx 1.18$.  We conjecture that
with increasing values of the phonon frequency the ``polaronic
instability'' is shifted to larger values of the electron phonon
coupling and in the shaded parameter regime of Fig. \ref{mass.fig}
strong isotope effects on electronic properties have to be expected.

If one includes the leading frequency correction to the finite loop
(ME) diagram, the effective mass reads\cite{deppeler02a}
\begin{equation}
\label{mass}
\frac{m^*}{m} = 1 + \bar{\Lambda}\,\rho(\mu) - 
\frac{1}{4}\,\bar{\lambda}\,\bar{\gamma}
\end{equation}
with $\bar{\gamma}$ being the renormalized adiabatic parameter
$\bar{\gamma} = \gamma\,\sqrt{1-\Lambda/\Lambda_c}$. Therefore, the
electronic effective mass is basically unchanged for phonon
frequencies in the adiabatic regime. Systematic adiabatic
(small-$\gamma$) expansions\cite{deppeler02a} at zero temperature give
all contributions of order $\bar{\gamma}$ and correct
Eq. (\ref{mass}). However, these contributions come with additional
factors $\bar{\lambda}^n$ ($n \geq 2$). Thus they are not small close
to the polaronic instability and an adiabatic expansion is not
possible. This is one reason why the semiclassical approach discussed
here is restricted to high to intermediate temperatures which help to
re-establish smaller lattice distortions.

The results of Benedetti and Zeyher\cite{benedetti98} (although at
finite temperatures) for $\gamma = 0.01$ are close to the classical
result and fall as they should in the shaded region of
Fig. \ref{mass.fig}. One of the authors has studied the consequences
of the isotope shift of $\Lambda_c$ when both thermal and quantum
fluctuations are competing by means of the semiclassical approach
revisited here. As one of the most dramatic manifestations an isotope
driven insulator-to-metal transition may be observed.\cite{blawid02}

\section{Adiabatic corrections}
\label{adiabatic}

From the above considerations it becomes obvious that adiabatic
corrections to the classical effective action have to be taken into
account. For the Holstein model this has been done in Ref.
\onlinecite{blawid02}. The idea is to expand the difference $S_{\rm
eff}-S_{\rm class}$ to second order in
$(1-\delta_{nm})\,\underline{\underline{W}}(\nu_n-\nu_m)$, i.e.~to
leading order in the non diagonal part of the electron-phonon
interaction. This yields 
\begin{equation} 
S_{\rm eff} = S_{\rm class} + S_2 
\end{equation} 
with 
\begin{eqnarray}
\label{adiabat} 
S_2 & = & \frac{1}{2T}\,
\sum_{k\neq 0} \vec{Q}(\omega_k) (K+M\omega_k^2) \vec{Q}(\omega_{-k}) +
\\ 
& & \frac{1}{2}\,\sum_{k\neq 0}\,{\rm Tr}_{\rm orb}\,
\left[c(\nu_n)-\underline{\underline{W}}(\vec{Q})\right]^{-1}
\underline{\underline{W}}[\vec{Q}(\omega_k)]
\nonumber \\
& & \times\,
\left[c(\nu_{n-k})-\underline{\underline{W}}(\vec{Q})\right]^{-1}
\underline{\underline{W}}[\vec{Q}(-\omega_k)] \nonumber 
\end{eqnarray}

In the following we will present numerical results for the Holstein
model at half-filling, i.e.~$\underline{\underline{W}} = g\,Q$ and
$\mu =0$. We generalize the treatment given in
Ref. \onlinecite{blawid02} by using a different numerical
implementation which does not rely on approximatively continuing the
local electronic Green function from the imaginary to the real
frequency axis. Instead, the continuation is done ``numerically''
exact. This allows for a more extensive discussion of the optical
conductivity.  We shortly review results from Ref.
\onlinecite{blawid02} pertinent to the present work.  For
$\underline{\underline{W}} = g\,Q$ all but one integration in Eq.
(\ref{partition}) can be performed analytically and the effective
action depends only on the classical coordinate $Q$
\begin{equation} 
S_{\rm eff}(Q) = S_{\rm class}(Q) +
\sum_{k>0} \ln \left[1+\frac{\Omega^2(\omega_k,Q)}{\omega_k^2}\right]\;.
\end{equation} 
Here, not the bare phonon frequency $\Omega^2 = K/M$ enters but a
renormalized one which depends on both the classical distortion $Q$
and the even Matsubara frequencies $\omega_k = 2\pi T\,k$ and reads
\begin{equation}
\label{om2}
\Omega^2(\omega_k,Q) = \Omega^2\left[1+\Pi(\omega_k,Q)\right] \;.
\end{equation}
The $\omega_k$ dependence arises from the fact that the phonon
propagator is renormalized by electron bubbles, i.e.~processes where
one electron absorbs (or emits) one phonon and gains (or loose) the
energy $\omega_k$. Therefore in Eq. (\ref{om2})
\begin{equation}
\Pi(\omega_k,Q) = \Lambda\,T\,\sum_{\nu_n} 
\frac{1}{c(\nu_n)-Q}\,\frac{1}{c(\nu_n-\omega_k)-Q}\;.
\end{equation}
If the renormalized phonon frequency were independent of the even
Matsubara frequencies $\omega_k = 2\pi T\,k$, the quantum term
$S_2$ were the action of a harmonic oscillator.  This
simplification was explored in Ref.\onlinecite{blawid02}. Here,
however, we will investigate the exact form of the second order
quantum corrections.

\begin{figure}
\centerline{\psfig{file=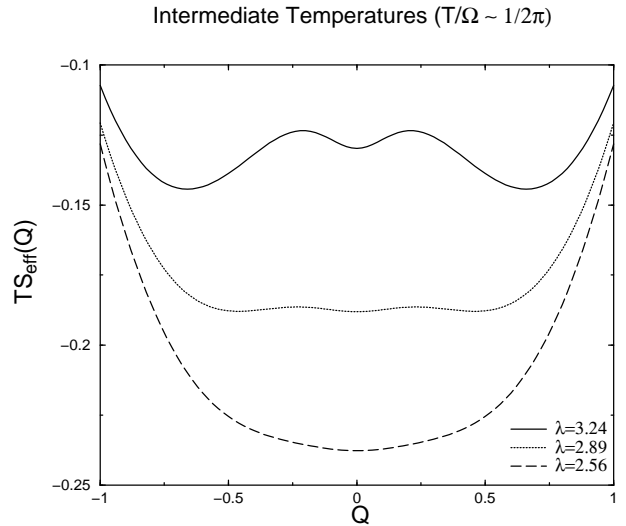,width=8cm,angle=-90}}
\vspace*{0.1cm}
\caption{\label{inter.fig} \footnotesize The effective action $S_{\rm
eff}$ as function of the distortion $Q$ for the Holstein model at
half-filling. The adiabatic parameter equals $\gamma = 0.2$. Shown is
the behavior for different coupling strengths $\lambda$ at
a intermediate temperature of $T/t=0.036$. The forming of a {\it
triple} (instead of a double) well at a finite value of $\lambda$ is
clearly seen.  }
\end{figure}
At weak couplings the mean-field functions $c(\nu_n)$ may be
approximated by the free ones. However, at intermediate to strong
electron phonon couplings the electronic properties are influenced and
the mean-field functions are renormalized as described by the
self-consistency equation (\ref{dmft}). The local Green function is
obtained by the logarithmic derivative of the partition function $Z$
with respect to $c(\nu_n)$, Eq. (\ref{greens}). It reads (denoting
$\frac{1}{Z}\,\int\,{\rm d}r\,P(Q)$ by $\langle\,\rangle$)
\begin{equation} 
\label{green} {\cal G}(\nu_n) = 
\left\langle \frac{1}{c(\nu_n)-Q} +
F(\nu_n,Q)\,\left(\frac{1}{c(\nu_n)-Q}\right)^2 \right\rangle\;.
\end{equation} 
Comparing to the classical result, the quantum corrections enter in
two places. First, they enter in the modified probabilty $P(Q) =
\exp(-S_{\rm eff})$ that the classic distortion takes the value $Q$.
Second, they give rise to an additional scattering term $F(\nu_n,Q)$
\begin{eqnarray} 
\label{full} 
F(\nu_n,Q) & = & \Lambda\,T\,
\sum_{k>0}\,\frac{\Omega^2}{w_k^2+\Omega_k^2(Q)} \\
& & \times\,
\left(\frac{1}{c(\nu_{n-k})-Q}+\frac{1}{c(\nu_{n+k})-Q}\right)
\nonumber
\end{eqnarray} 
which can be interpreted as generalized Migdal self energy where the
classical $k = 0$ term is subtracted. The two modifications have
opposing influences. First, as shown in Fig.\ref{inter.fig}, the
adiabatic corrections lead to an increased propability for distortions
around $Q=0$ compared to the classical one. In consequence, quantum
fluctuations of the lattice (not present in the classical approach)
may lead to an increased mobility of the conduction electrons. Second,
when an analytic continuation to the real frequency axis of $F$ is
performed [yielding $F(\omega,Q)$], a finite imaginary part implies
additional electron phonon scattering. Note, that for small
frequencies $\omega$ below the renormalized one the imaginary part of
$F(\omega,Q)$ may be reduced or vanishing.

\subsection{Numerical implementation}

Iteration of Eq. (\ref{dmft}) and Eq. (\ref{green}) will yield the
local electronic Green function ${\cal G}(\nu_n)$ on the imaginary
axis.  The occuring fermionic and bosonic Matsubara sums are
calculated introducing high frequency cut offs $\bar{\omega}$ and
$\bar{\nu}$, respectively. The high frequency contributions to the
sums are approximated by integrals. For reference we summarize here
the results where functions $f(\bar{\omega})$ [$f(\bar{\nu})$] are
finite bosonic (fermionic) Matsubara sums up to $\omega_k <
\bar{\omega}$ (up to $\nu_n < \bar{\nu}$):

\begin{eqnarray}
S_{2}(Q) & = & S_{2}(Q,\bar{\omega}) +
\frac{\Omega}{2\,T} -
\frac{2 \Omega}{2 \pi T} \, \arctan \frac{\bar{\omega}}{\Omega} \\
& & -
\frac{\bar{\omega}}{2 \pi T} \, 
\ln \left(1+\frac{\Omega^2}{\bar{\omega}^2}\right)
\nonumber\\
\Pi(\omega_k,Q) & = & \Pi(\omega_k,Q,\bar{\nu}) \\
& & -
\frac{1}{2\,\pi\,\omega_k}\,\ln 
\left( 
1 + \frac{2\,\bar{\nu}\,\omega_k+\omega_k^2}{Q^2+\bar{\nu}^2}
\right)
\nonumber\\
\label{scattasymp}
F(\nu_n,Q) & = & F(\nu_n,Q,\bar{\omega}) +
\frac{\Lambda\,\Omega}{\pi}\,\frac{1}{Q^2-\Omega^2}\\
& & \times\,\left[
Q\,(\arctan\frac{\bar{\omega}}{\Omega}-\frac{\pi}{2})
\right. \nonumber\\
& &  - \left.
\Omega\,(\arctan\frac{\bar{\omega}}{Q}-{\rm sgn}Q\,\frac{\pi}{2})
\right]\;.
\nonumber
\end{eqnarray}

In principle, the truncation at $\bar{\omega}$ and the adding of an
asymptotic contribution in Eq. (\ref{scattasymp}) is not valid if $\nu_n
\approx \bar{\omega}$, i.e.~for large values of the fermionic
Matsubara frequency. However, the quantum scattering term $F(\nu_n,Q)$
tends to zero in this case and small errors are not important. In all
our calculations we used 128 bosonic (fermionic) Matsubara
frequencies. To speed up the integration performed in Eq. (\ref{green})
the electron electron bubble $\Pi(\omega_k,Q)$ has been interpolated
as function of $Q$ for each value of $\omega_k$ in each DMFT iteration
step.

In the present approach only one dimensional integrals over a real
variable have to be calculated. This can be done with high precision
yielding ${\cal G}(\nu_n)$ up to 14 digits. Therefore, we can continue
the electronic Green function to the real frequency axis employing a
``simple'' Pad{\'e} approximation.\cite{vidberg77} The quality of the
results are even more improved when using the classical solution as a
reference state, i.e. doing the continuation for the difference ${\cal
G}-{\cal G}_{\rm class}$. This is a real strength of the present
semiclassical approach allowing the calculation of dynamic quantities
without further assumptions. As important example we have calculated
here the finite temperature optical conductivity of the Holstein model
as convolution of two full Green functions:
\begin{eqnarray}
\label{cond}
\sigma(\Omega) & = & \sigma_0\,
\int {\rm d}\omega\,
\left(-\frac{f(\omega+\Omega)-f(\omega)}{\Omega}\right) \\
& & \times\,
\int {\rm d}\epsilon\,
\rho_0(\epsilon)\,\rho(\epsilon,\omega)\,\rho(\epsilon,\omega+\Omega)\;.
\nonumber
\end{eqnarray}
The external frequency $\Omega$ shall not be confused with the phonon
frequency. In Eq. (\ref{cond}) is $\rho_0(\epsilon)$ of semicircular
form and the spectral function defined by
\begin{equation}
\rho \left( \epsilon_{\vec{k}},\omega \right)  =  
-\frac{1}{\pi}\,{\rm Im} 
\left[
\frac{1}{\omega+\mu-\Sigma(\omega)-\epsilon_{\vec{k}}}
\right]\;.
\end{equation}
The self-energy $\Sigma(\omega)$ can be obtained from the local
electronic Green function ${\cal G}(\omega)$ on the real frequency
axis. We also present results for the density of states
\begin{equation}
\rho(\omega) =  \int {\rm d}\epsilon\,
\rho_0(\epsilon)\,\rho(\epsilon,\omega)\;. 
\end{equation}

To summarize, our implementation of the semicalssical approach for the
Holstein model differs in two way from the one discussed in
Ref. \onlinecite{blawid02}. First, when calculating the effective
action Eq. (\ref{adiabat}) we do not rely on a static approximation,
i.e.~we do not replace $\Omega(\omega_k,Q) \rightarrow
\Omega(0,Q)$. However, this has little influence on the
results. Second and most important, ${\cal G}(\nu_n)$ is continued to
the real axis not approximatively but numerically exact. In
particular, the investigations of dynamic quantities for
electron-phonon couplings above the (classical) polaronic instability
is possible which is not achieved in Ref. \onlinecite{blawid02}.

\subsection{Limitations}

Before presenting results we shortly summarize the shortcommings of
the presented semiclassical approach as discussed extensively in
Ref.\onlinecite{blawid02}. Within the framework of dynamical mean-field
theory the approach becomes exact at high temperatures $T \rightarrow
\infty$, irrespectively which value the electron phonon coupling
takes. However, at low temperatures the approach fails for two
reasons:

First, at intermediate to strong couplings the weak coupling (Fermi
liquid) ground state is fundamentally reconstructed. In the present
formalism this is reflected by the occurence of imaginary phonon
frequencies $\Omega^2(\omega_k,Q)$. In this case $S_2(Q)$ is not well
defined anymore at low temperatures. Only larger temperatures
re-stabilize the ground state. To explore this in more detail, let us
suppose the renormalized phonon frequency $|\Omega(\omega_k=0,Q=0)|$
turns imaginary at a critical coupling $\Lambda_c$. For $ T \rightarrow
0$ and for couplings just above the critical value $\Delta\Lambda =
\Lambda-\Lambda_c$ one finds $\Omega^2(0,0) =
-\Omega^2\,\Delta\Lambda/\Lambda_c$.\cite{blawid02} Because
$\Omega^2(\omega_k,Q)$ is a strictly monotonic decreasing function
with Matsubara frequency and displacement, the quantum correction
$S_2(Q)$ is only defined if
\begin{equation}
T > \frac{\Omega}{2\,\pi} \left|\Omega(0,0)\right| = 
 \frac{\Omega}{2\,\pi} \sqrt{\frac{\Delta\Lambda}{\Lambda_c}}\;.
\end{equation}
However, at high temperatures $\Omega^2(0,0) = \Omega^2 - \Lambda/(4
T)$ and the renormalized phonon frequency is always real.

Second, also at weak couplings the ground state cannot be reached. A
purely classical approximation ($\Omega \rightarrow 0$) would lead to
a $T$-linear scattering rate with coefficient of order unity. The
semiclassical approximation overcorrects for this behavior, cancelling
the leading term leaving a correction of order $\gamma T$ but of the
wrong sign. Note that for $T < \Omega$ this is ${\cal O}(\gamma^2)$.
Note also that at $T = 0$ the adiabatic expansion is possbile for weak
couplings (see Sec. \ref{quantum} and Ref. \onlinecite{deppeler02a}).

In conclusion the semicalssical approach is restricted to a range of
high ($ T/\Omega \gg 1/2 \pi$) to intermediate ($T/\Omega \sim 1/2
\pi$) temperatures and we present only results in this range.

\subsection{Results}

In the following we present numerical results for the density of
states (DOS) and the optical conductivity at finite temperatures for
electron-phonon couplings from weak to strong. The model under
consideration is the Holstein model with one (spinless) electron per
each two sites (half filling), i.e.~we discuss the simpliest case of
non-degenerate electrons coupling linearly to non-degenerate lattice
distortions.

\begin{figure}
\centerline{\psfig{file=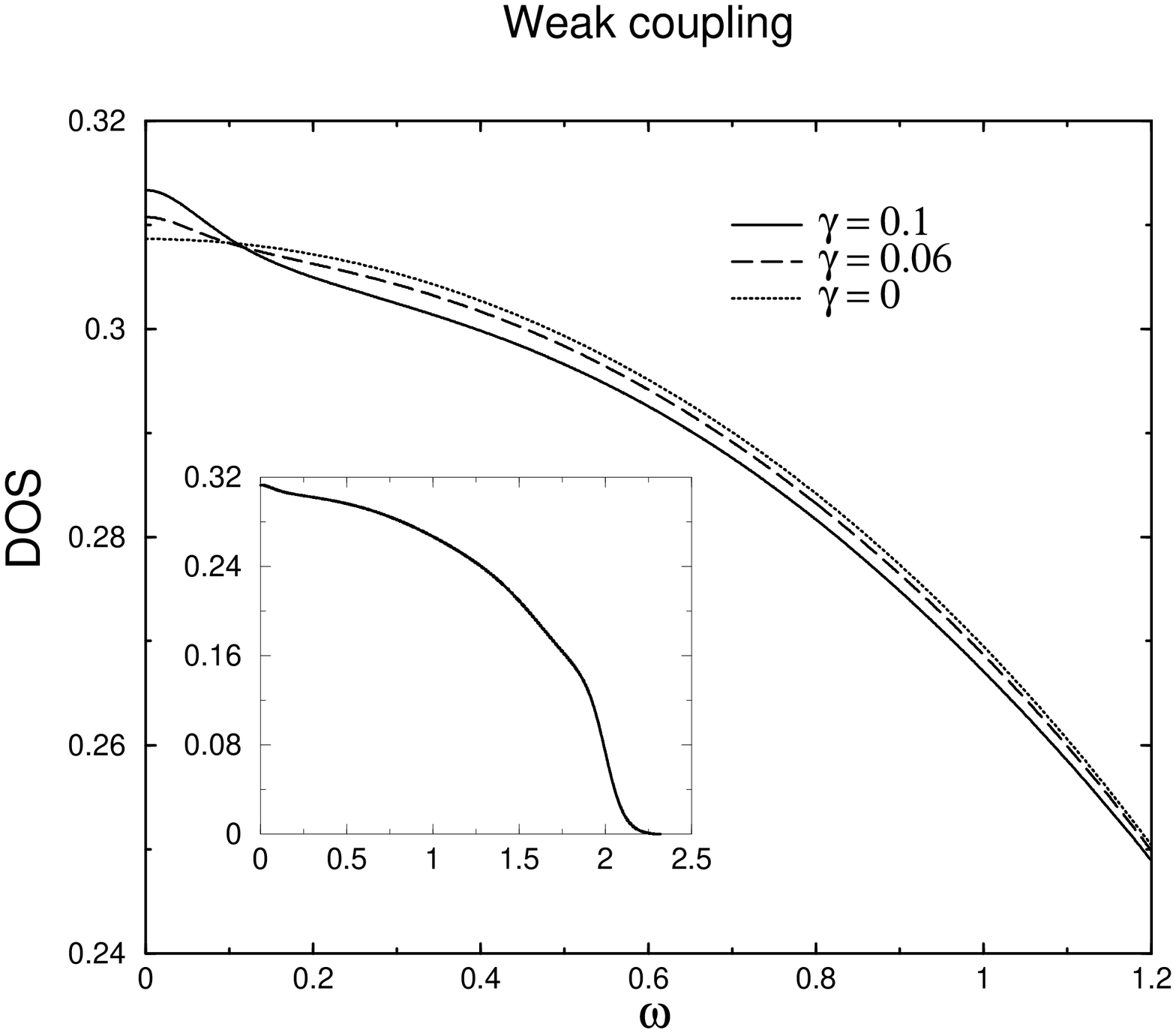,width=8cm,angle=-90}}
\centerline{\psfig{file=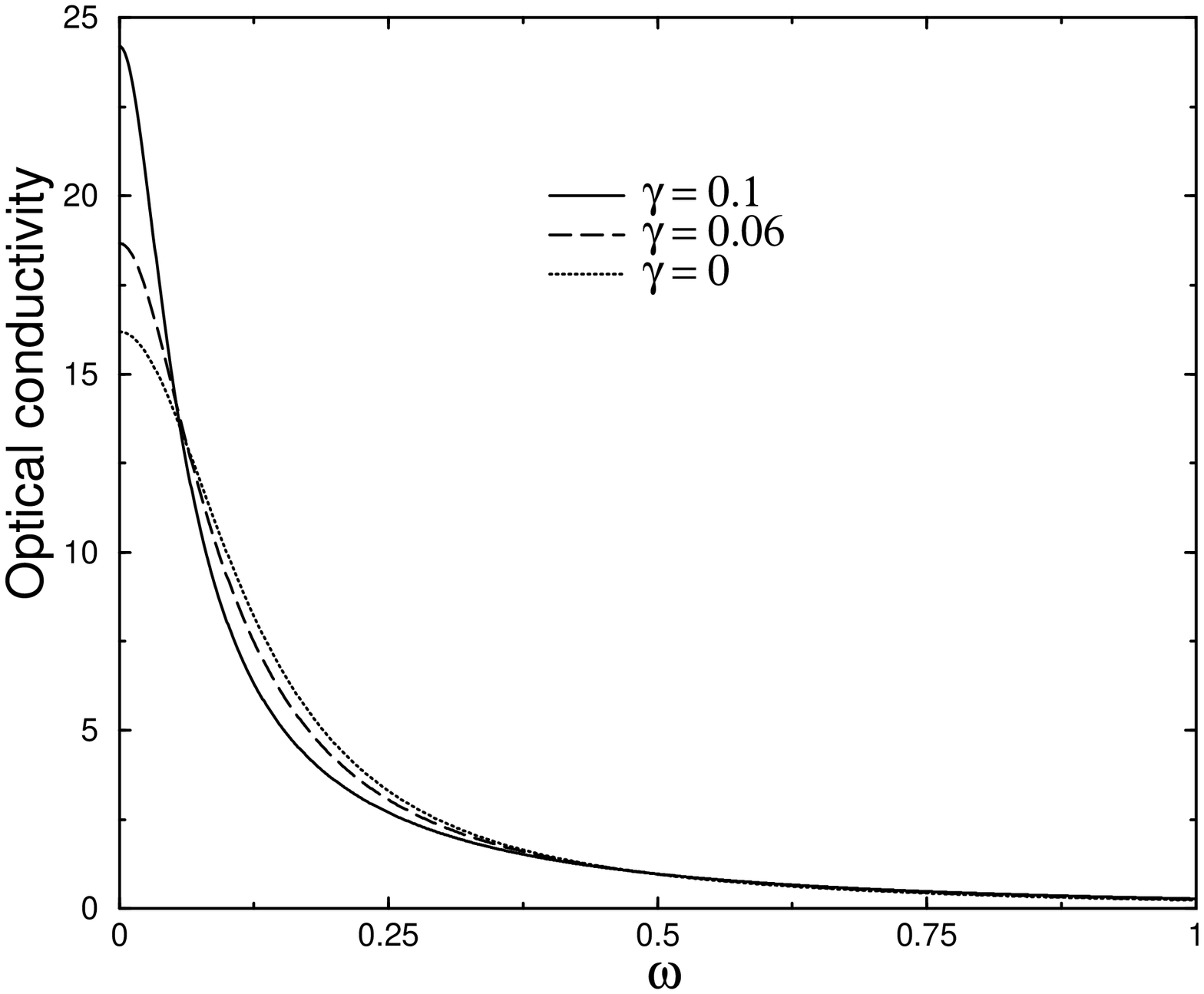,width=8cm,angle=-90}}
\vspace*{0.1cm}
\caption{\label{dynweak.fig} \footnotesize UPPER PANEL: DOS of the
Holstein model (half-filling) at a {\it weak} electron-phonon coupling
$\lambda = 1$ and a temperature $T/t = 0.04$. Results for various
values of the adiabatic parameter $\gamma$ are shown. For
finite phonon frequencies a small quasi-particle peak is visible at
$\omega = 0$. The inset displays the DOS over the full frequency
range for the largest value of the adiabatic parameter, $\gamma =
0.1$. On this scale the quasiparticle peak is hardly seen. LOWER
PANEL: The frequency dependend optical conductivity for the same model
and parameters as in the upper panel. It displays a Drude-like
behavior.}
\end{figure}
In Fig. \ref{dynweak.fig} we show typical results at {\it weak}
couplings, i.e. for values of $\Lambda$ well below the (classical)
polaronic instability of $\Lambda_c \approx 2.36$. We recover the
behavior known from calculations employing Migdal-Eliashberg
theory.\cite{hague01} A tiny quasi-particle peak at zero frequency
(and the inset in the upper panel showing the DOS over the complete
frequency range shall demonstrate that it is indeed a tiny structure)
is observed in the DOS and {\it no} upper and lower sub-bands are
formed. The quasi-particle peak originates from the fact that
conductions electrons can only inelastically scatter from quantum
lattice fluctuations with a finite frequency transfer. Therefore, the
peak gets slightly broader with increasing adiabatic parameter
$\gamma$. Note that in all our calculations $\gamma$ is a small
parameter. Following the behavior of the DOS, the optical conductivity
shows a typical metallic behavior with a sharp Drude peak. The optical
conductivity drops monotonically with increasing applied frequency
$\omega$.

\begin{figure}
\centerline{\psfig{file=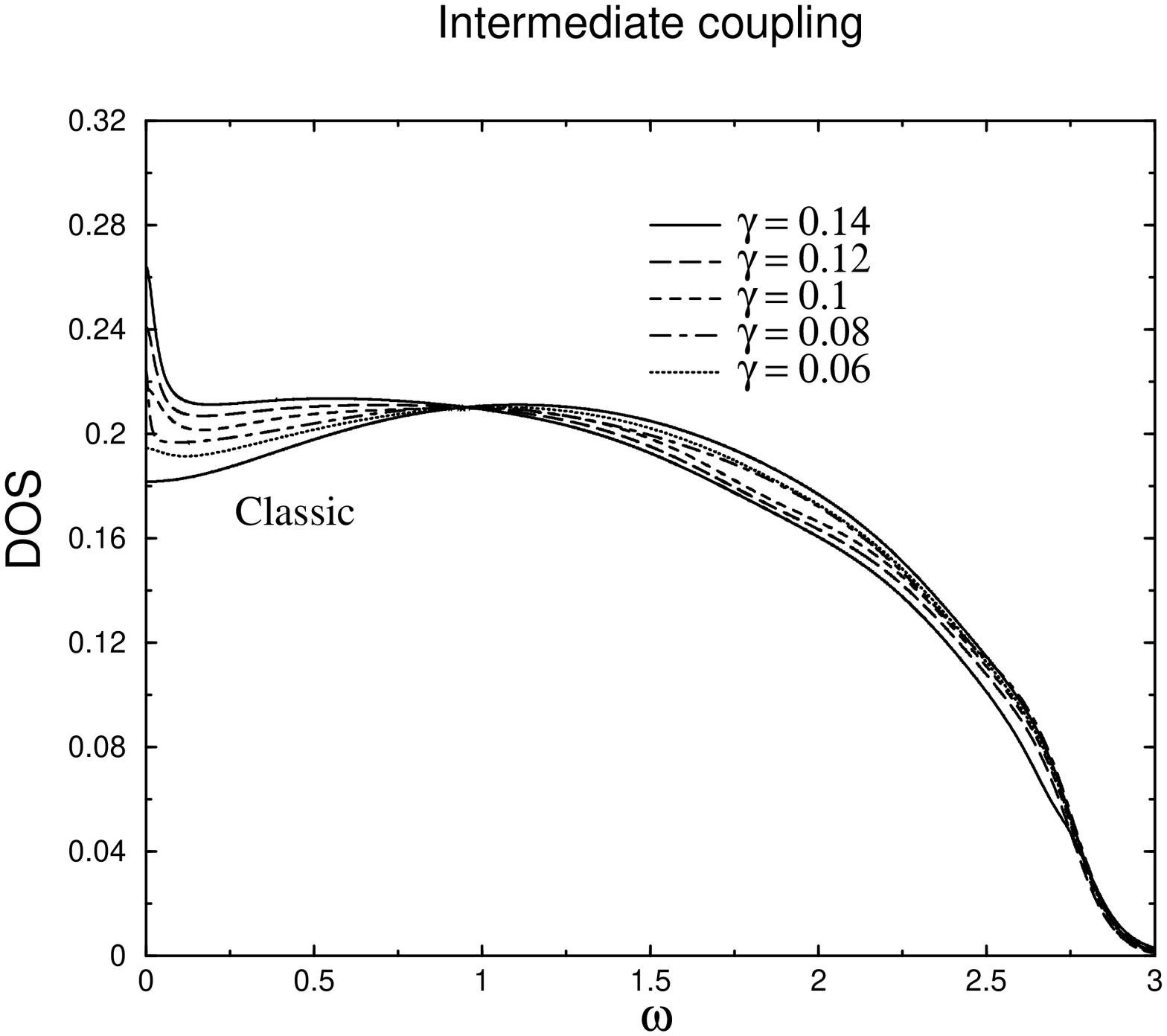,width=8cm,angle=-90}}
\centerline{\psfig{file=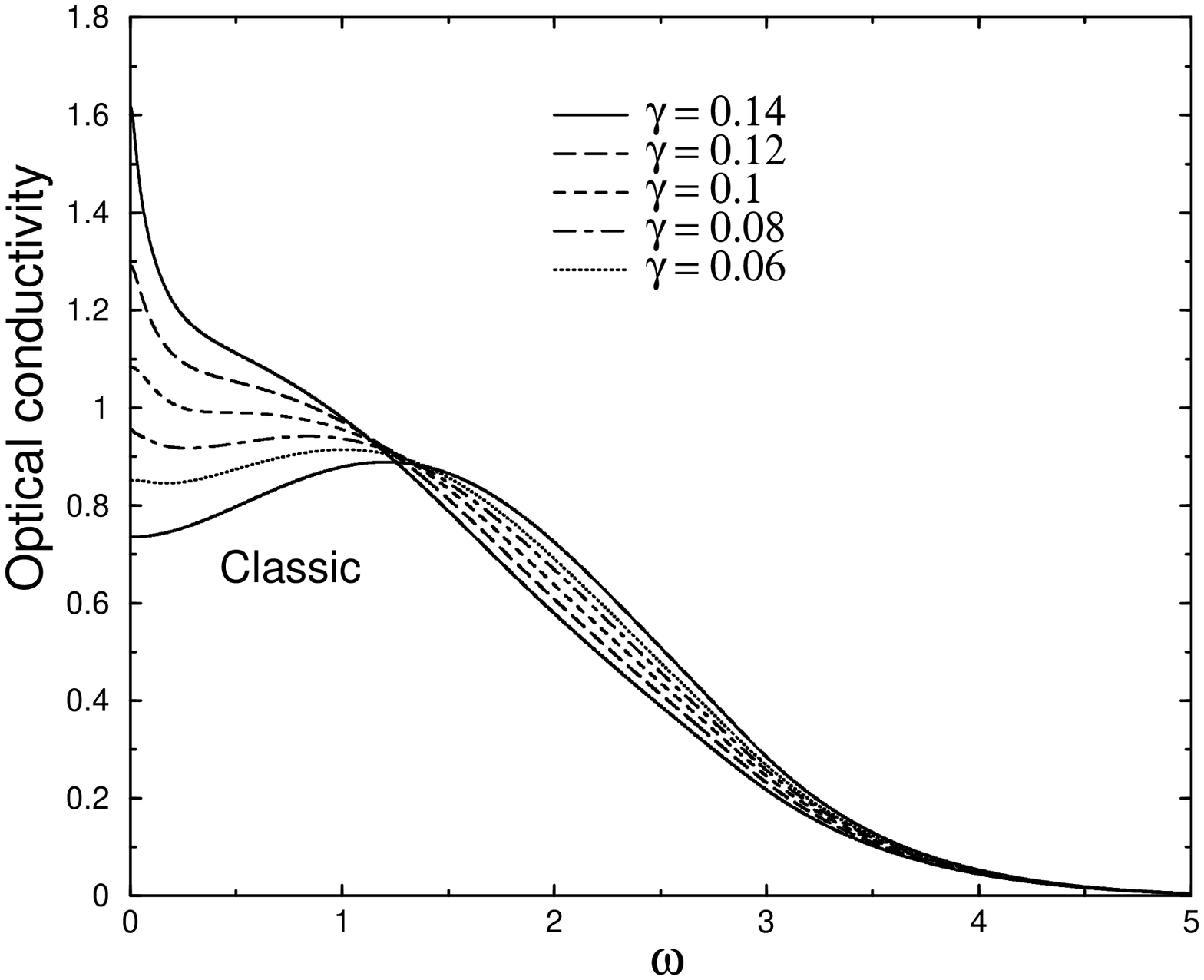,width=8cm,angle=-90}}
\vspace*{0.1cm}
\caption{\label{dyninter.fig} \footnotesize UPPER PANEL: DOS of the
Holstein model (half-filling) at an {\it intermediate} electron-phonon
coupling $\lambda = 2.89$ and a temperature $T/t = 0.04$. Results for
various values of the adiabatic parameter $\gamma$ are
shown. Increasing the adiabatic parameter has a considerable
influence.  Weakly developed sub-bands occuring at small phonon
frequencies disappear in favour of a narrow quasi-particle peak at
$\omega = 0$ with increasing $\gamma$. LOWER PANEL: The frequency
dependend optical conductivity for the same model and parameters as in
the upper panel. At small phonon frequencies an intra-band absorption
is clearly visible. With increasing phonon frequency this feature
merges with a developing Drude peak giving rise to a ``shoulder''.}
\end{figure}
Fig. \ref{dyninter.fig} shows the DOS and the optical conductivity at
an {\it intermediate} electron-phonon coupling, just above the
(classical) polaronic instability. In this regime isotope effects
(i.e.~changing values of $\gamma$) may alter the behavior {\it
qualitatively}. At low phonon frequencies sub-bands in the DOS are
weakly developed. With increasing $\gamma$ these sub-bands disappear
in favor of a quasi-particle peak at $\omega = 0$. The
quasi-particle peak is more pronounced as in the weak coupling case
with slightly smaller width. With increasing adiabatic parameter also
the behavior of the optical conductivity changes qualitatively from an
insulating behavior with increased {\it intra-band} absorption to a
Drude-like metallic behavior with an {\it incoherent contribution}
visible as ``shoulder''. Although a Drude peak develops the conductivity
is one order of magnitude smaller than for small electron-phonon
couplings. One may speak of a ``bad metal''.

\begin{figure}
\centerline{\psfig{file=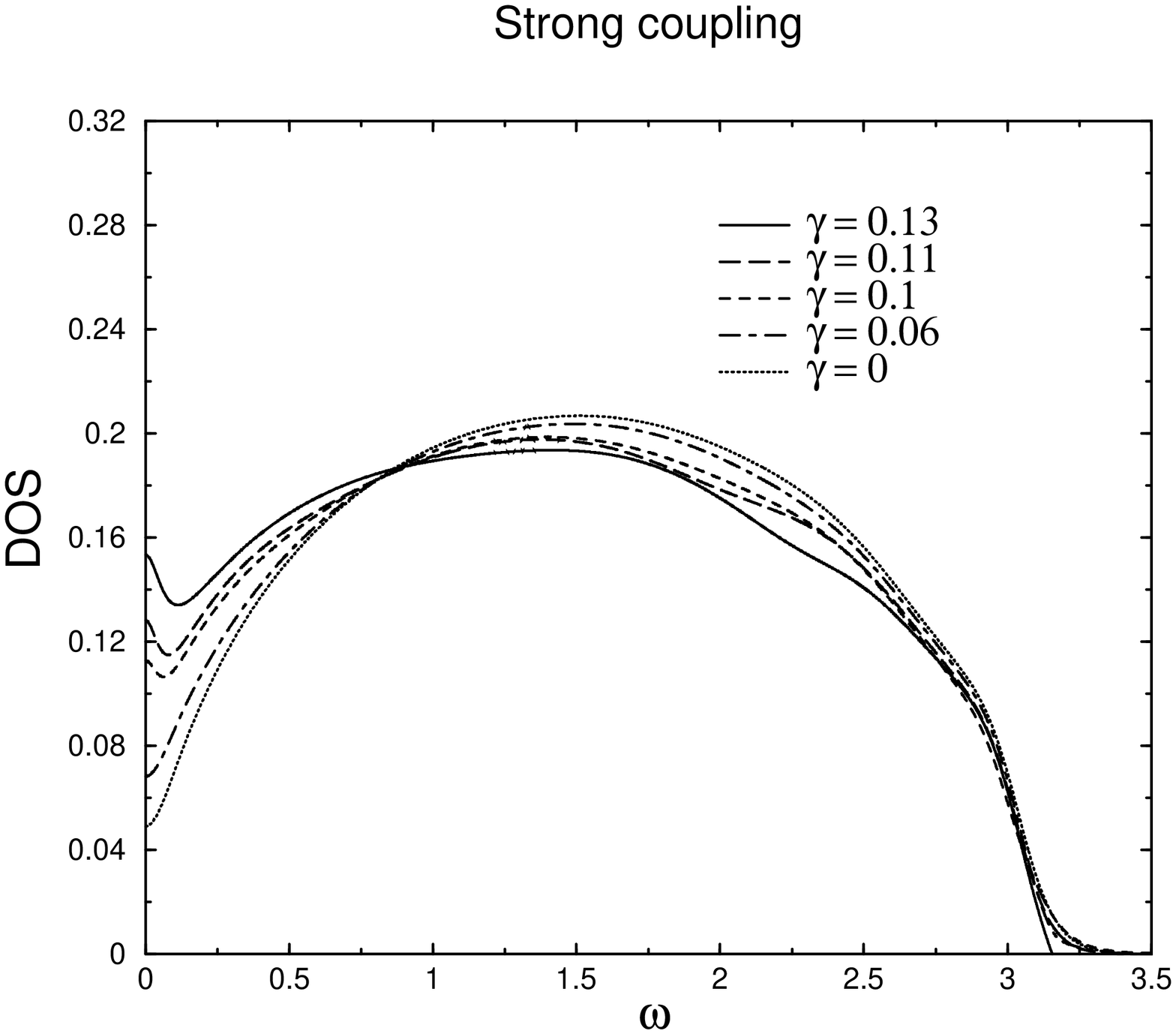,width=8cm,angle=-90}}
\centerline{\psfig{file=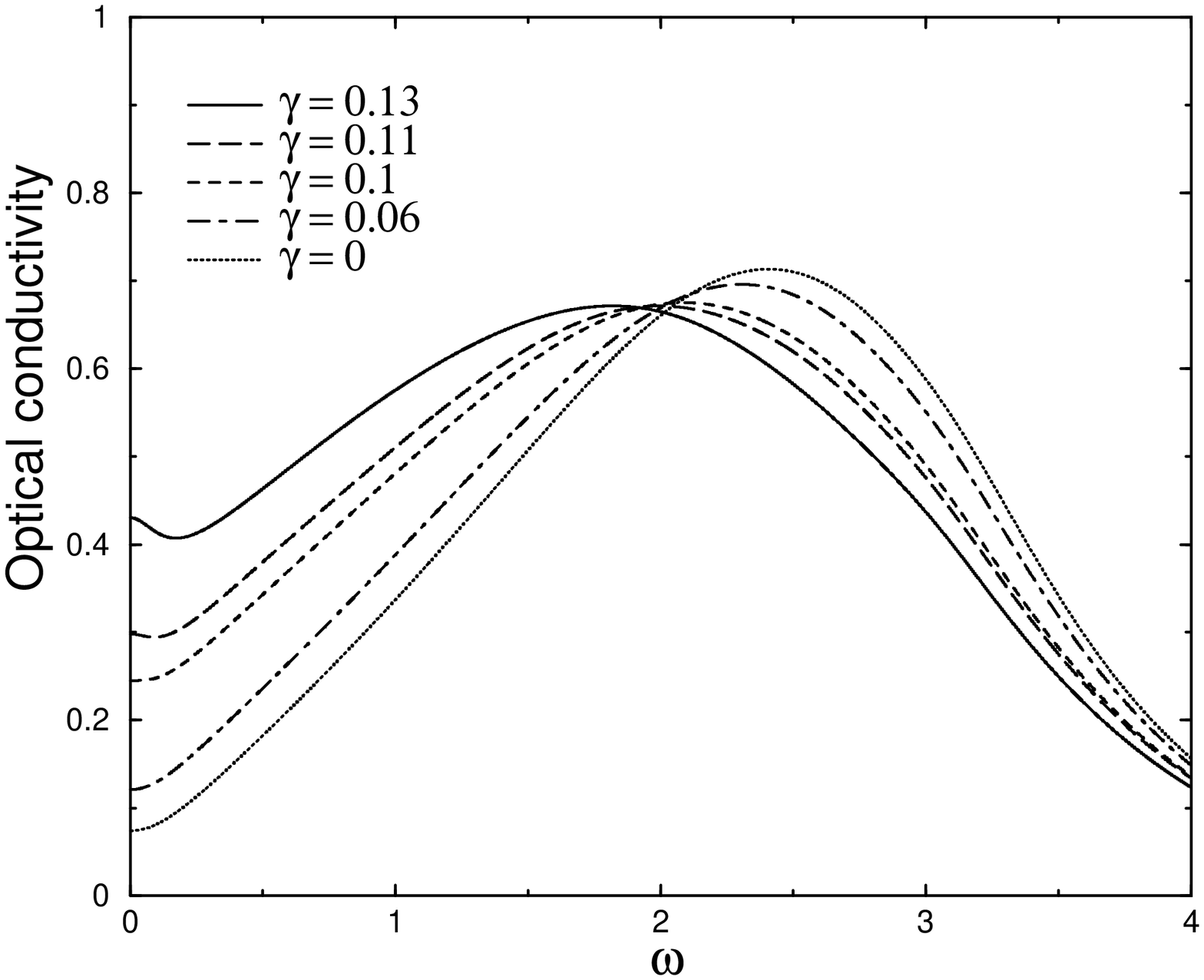,width=8cm,angle=-90}}
\vspace*{0.1cm}
\caption{\label{dynstrong.fig} UPPER PANEL: DOS of the Holstein model
(half-filling) at a {\it strong} electron-phonon coupling $\lambda =
3.24$ and a temperature $T/t = 0.04$. Results for various values of
the adiabatic parameter $\gamma$ are shown. For all values of $\gamma$
the DOS displays well developed sub-bands. A quasi-particle peak
becomes visible for larger values of the phonon frequency. LOWER
PANEL: The frequency dependend optical conductivity for the same model
and parameters as in the upper panel. An asymmetric inter-band
absorption is clearly seen situated roughly at $\omega \approx
\Lambda$, shifting towards smaller frequencies with increasing
$\gamma$. }
\end{figure}
Results for {\it stronger} couplings are shown in
Fig. \ref{dynstrong.fig}. Sub-bands are well developed in both the
DOS and the optical conductivity. Only for a significant change in the
phonon frequency a Drude-like metallic behavior is restored. In this
sense isotope effects are less pronounced in the strong coupling case
as in the intermediate one as they do not lead to immediate
qualitative changes. The qualtitative behavior as given in the
classical picture is maintained for all small values of the adiabatic
parameter $\gamma$. However, at larger phonon frequencies both a
quasi-particle peak at $\omega = 0$ and sub-bands are visible
simultanously, a feature which cannot be obtained within
Migdal-Eliashberg theory. The intra-band absorption in the optical
conductivity occurs around $\omega \approx \Lambda$ and shifts to
smaller frequencies with increasing value of the adiabatic parameter.
The intra-band absorption is asymmetric about its peak. In particular,
the low-frequency side of the absorption peak is elevated above the
high-frequency side.

The low temperature behavior of the DOS and the optical conductivity
in the adiabatic regime has been carefully studied by Benedetti and
Zeyher.\cite{benedetti98} Their results confirm nicely that our
findings at intermediate temperatures are not spoiled by the low
temperature break down of the semiclassical approach but that they
reflect the physical behavior of the model. In particular, the optical
conductivity shows the behavior presented here: (i) Drude like
behavior at weak coupling, (ii) quasi-particle peak together with
incoherent features at intermediate couplings and (iii) intra-band
absorption at strong coupling. Obviously, the present semiclassical
approach is capable of describing the interesting crossover regime
from high to low temperatures. Unfortunately, isotope effects on the
optical conductivity were not discussed in
Ref.\onlinecite{benedetti98} which are especially relevant when
Migdal-Eliashberg theory breaks down (i.e. at intermediate
electron-phonon coupling strengths).

It should be noted that the qualitative features and changes in the
optical conductivity when increasing the electron-phonon coupling from
weak to strong are similar to the ones predicted by a {\it
small-polaron} absorption theory.\cite{emin93} However, in this theory
the hopping of the conduction electrons is a small parameter contrary
to the approach discussed here which is based on an adiabatic
expansion. Therefore a comparison is only meaningfull in the limit of
large electron-phonon couplings where small polarons are well defined
quasi-particles. Here, the main absorption process is the excitation
of a self-trapped electron from its localized state at one site to a
localized state at a neighboring one. Due to thermal and quantum
fluctuations of the lattice the energies of the localized states are
broadened and one obtains for the optical conductivity
\begin{equation}
\sigma \sim \frac{1}{\omega}\,
\exp[-(\Lambda-\omega)^2/(4\,\Lambda\,E_{\rm vib})]
\end{equation}
where $E_{\rm vib} = \Omega/2$ is the zero-point vibrational energy at
low temperatures and $E_{\rm vib} = T$ in the case of high enough
temperatures when the vibrations can be treated classically. The
classical (high temperature) result can be compared to our numerical
datas in the strong coupling limit and reasonable agreement is found.

Depending on the values of the phonon frequency $\Omega$ and coupling
strength $\Lambda$ qualitative features of various experiments are
present. To name two examples: (i) In $\rm Ba_{1-x}K_xBiO_3$ for high
enough doping $x$ (so that charge ordering is destroyed) the optical
conductivity displays a Drude-like peak with a pronounced incoherent
feature in the mid infrared;\cite{puchkov96} (ii) Asymmetric
intra-band absorption can be observed in the alkali doped fullerides
$\rm K_3C_{60}$ and $\rm Rb_3C_{60}$.\cite{degiorgi94} As outlined in
this paper the semiclassical approach can be extended to more
realistic models of electrons in degenerate orbital states coupling to
Jahn-Teller distortions allowing for a more detailed analysis of
experiments. Along the lines of Refs. \onlinecite{deppeler00} and
\onlinecite{deppeler02b} effects of {\it local} Coulomb repulsion may
be included.

\section{Conclusion}
\label{conclusion}

In this paper we have extended a semiclassical approach\cite{blawid02}
based on the dynamical mean-field theory recently introduced by one of
the authors to treat the coupling of conduction electrons to local
Jahn-Teller distortions in metals and metallic molecular crystals. We
have improved previous numerical treatments allowing a detailed
discussion of dynamic quantities such as the finite temperature
optical conductivity and presented results for the simplifying
Holstein model. In particular, the density of states and the optical
conductivity could be studied not only for electron-phonon couplings
below the (classical) polaronic instability but also above. At
intermediate to strong couplings the optical conductivity displays
features which are known from experiments on phonon mediated
superconductors with high transition temperatures. Depending on
paramters, a Drude-like peak with a pronounced incoherent ``shoulder''
or an asymmetric intra-band absorption is observed. Strong isotope
effects occur at couplings close to the polaronic instability changing
the qualitative behavior of the system.
 
The presented results demonstrate that the analysis of simplified
effective actions based on the dynamical mean-field theory and
adiabatic expansions is a powerful tool to study the influence of
local lattice fluctuations on electronic properties of metals. Future
work will reveal if the presented approach is capable also of a {\it
quantitative} analysis of electronic properties of metals and metallic
molecular crystals strongly influenced by local Jahn-Teller
distortions.

{\it Acknowledgements} S.B.~thanks A.~Millis for many usefull and
stimulating discussions. S.B.~acknowledges the Niedersachsen-Israel
foundation for financial support at early stages of this work.



\vspace*{-6mm}

\begin{references}

\vspace*{-14mm}

\bibitem{puchkov96} A.V.~Puchkov, T.~Timusk, M.A.~Karlow, S.L.~Cooper,
P.D.~Han and D.A.~Payne, Phys. Rev. B {\bf 54}, 6686 (1996)

\bibitem{degiorgi94} L.~Degiorgi, E.J.~Nicol, O.~Klein, G.~Gr\"uner,
P.~Wachter, S.M.~Huang, J.~Wiley and R.B.~Kaner, Phys. Rev. B {\bf
49}, 7012 (1994)

\bibitem{gunnarsson97} O.~Gunnarsson, 
Rev.~Mod.~Phys. {\bf 69}, 575 (1997)

\bibitem{migdal58}
A.B.~Migdal, Sov. Phys. JETP {\bf 7}, 996 (1958); 
G.M.~Eliashberg, Sov. Phys. JETP {\bf 11}, 696 (1960)

\bibitem{alexandrov94} A.S.~Alexandrov, V.V.~Kabanov and D.K.~Ray,
Phys. Rev. B {\bf 49}, 9915 (1994)

\bibitem{lang63} I.G.~Lang and Yu.A.~Firsov, Sov. Phys. JETP {\bf 16},
1301 (1963)

\bibitem{georges96} A.~Georges, G.~Kotliar, W.~Krauth and
M.J.~Rozenberg, Rev. Mod. Phys. {\bf 68}, 13 (1996)

\bibitem{ciuchi96} S.~Ciuchi, F.~de~Pasquale, S.~Fratini and D.~Feinberg,
Phys. Rev. B {\bf 56}, 4494 (1997)

\bibitem{freericks}
J.K.~Freericks, M.~Jarrell, D.J.~Scalapino, Phys. Rev. B {\bf 48},
6302 (1993); J.K.~Freericks and M.~Jarell, Phys. Rev. B {\bf 50}, 6939
(1994); J.K.~Freericks, V.~Zlati{\'c}, W.~Chung and M.~Jarrell,
Phys. Rev. B {\bf 58}, 11613 (1998) 

\bibitem{meyer02} 
D.~Meyer, A:C:~Hewson and R.~Bulla, 
Phys. Rev. Lett. {\bf 89}, 196401 (2002)

\bibitem{millis96} A.J.~Millis, R.~Mueller and B.I.~Shraiman,
Phys. Rev. B {\bf 54}, 5389 (1996). A.J.~Millis, R.~Mueller and
B.I.~Shraiman, Phys. Rev. B {\bf 54}, 5405 (1996)

\bibitem{benedetti98} P.~Benedetti and R.~Zeyher, Phys. Rev. B, {\bf
58}, 14320 (1998)

\bibitem{deppeler00} A.~Deppeler and A.J.~Millis,
Phys. Rev. B {\bf 65}, 100301 (2002)

\bibitem{blawid02}
S.~Blawid, A.~Deppeler and A.J.~Millis,
Phys. Rev. B {\bf 67}, 165105 (2003)

\bibitem{deppeler02a}
A.~Deppeler and A.J.~Millis,
Phys. Rev. B {\bf 65}, 224301 (2002)

\bibitem{obrien97} C.C.~Chancey and M.C.M.~O'Brien, {\it The
Jahn-Teller effect in $\rm C_{60}$ and other icosahedral complexes}
(Princeton University Press, 1997)

\bibitem{vidberg77} H.J.~Vidberg and J.W.~Serene, J. Low
Temp. Phys. {\bf 29}, 179 (1977)

\bibitem{hague01} J.P.~Hague and N.~d'Ambrumenil,
cond-mat/0106355 (unpublished)

\bibitem{emin93} D.~Emin, Phys. Rev. B {\bf 48}, 13691 (1993)

\bibitem{deppeler02b} A.~Deppeler and A.J.~Millis, 
cond-mat/0204617

\end{references}
\end{document}